\documentclass[9pt,twocolumn,twoside]{opticajnl}
\journal{opticajournal} % use for journal or Optica Open submissions

\setboolean{shortarticle}{true}
\title{Enhancing the coherence time of a neutral atom by an optical quartic trap}

\author[1,2]{Haobo Chang}
\author[1,2]{Zhuangzhuang Tian}
\author[1,2]{Xin Lv}
\author[1,2]{Mengna Yang}
\author[1,2]{Zhihui Wang}
\author[1,2,3]{Qi Guo}
\author[1,2]{Pengfei Yang}
\author[1,2]{Pengfei Zhang}
\author[1,2,*]{Gang Li}
\author[1,2,$\dagger$]{Tiancai Zhang}

\affil[1]{State Key Laboratory of Quantum Optics Technologies and Devices, and Institute of Opto-Electronics, Shanxi University, Taiyuan 030006, China}
\affil[2]{Collaborative Innovation Center of Extreme Optics, Shanxi University, Taiyuan 030006, China}
\affil[3]{College of Physics and Electronic Engineering, Shanxi University, Taiyuan 030006, China}

\affil[*]{gangli@sxu.edu.cn}
\affil[$\dagger$]{tczhang@sxu.edu.cn}

\begin{abstract}
The coherence time of an optically trapped neutral atom is a crucial parameter for quantum technologies. We found that optical dipole traps with higher-order spatial forms inherently offer lower decoherence rates compared to those with lower-order spatial forms. We formulated the decoherence rate caused by the variance of the differential energy shift and photon jumping rate. Then, we constructed blue-detuned harmonic and quartic optical dipole traps, and experimentally investigated the coherence time of a trapped single cesium atom. The experimental results qualitatively verified our theory. Our approach provides a novel method to enhance the coherence time of optically trapped neutral atoms.

\end{abstract}

\setboolean{displaycopyright}{false} % Do not include copyright or licensing information in submission.

\begin{document}

\maketitle

As one of the basic parameters for quantum technologies, a long coherence time is always a stringent requirement.
In the context of optically trapped neutral atoms, research on coherence time (usually between two electronic states in the ground energy level) has garnered intense attention in the past decades \cite{03Kuhr,2005diter,  Jai2007, Kim2013, Derevianko2010, Derevianko2011, Kazakov2015, Yang2016, Guo2020, Carr2016, Li2019, Sarkany2014, Chicireanu2011, PFYang_2022, Tian2024}. 
Two primary decoherence mechanisms have been identified, and various schemes have been proposed to mitigate them.
One decoherence mechanism is the variance of the differential energy shift (DES) between the two qubit states due to the noises associated with the optical trapping and magnetic fields \cite{03Kuhr, 2005diter}. 
Many ``magic trapping conditions'', by which the variance of DES is immune to the external noises, have been discovered to suppress the DES-related decoherence, such as magic polarization \cite{Jai2007, Kim2013}, magic wavelength \cite{Derevianko2010, Derevianko2011, Kazakov2015, Li2019}, magic intensity \cite{Yang2016, Carr2016, Guo2020, Li2019} and magic magnetic field \cite{Carr2016, Chicireanu2011, Sarkany2014, Derevianko2010, Derevianko2011, Li2019}, etc.
The other mechanism of decoherence is the phonon-jumping-induced decoherence (PJID) \cite{Tian2024}, in which stochastic jumping of the atomic phonon state in an optical dipole trap can cause decoherence between the electronic states of the atoms.
The variance of DES and the PJID can be suppressed by taking a blue-detuned optical dipole trap and cooling the trapped single atom to its three-dimensional ground phonon states \cite{Tian2024, 19Weiss_SGdetection}.
The preceding works concerned the suppression of atom decoherence in optical traps with harmonic potential.
In this Letter, we study the dependence of the decoherence mechanism on the spatial form of the optical dipole trap. 
We found that a dipole trap with a higher order of spatial form possesses intrinsic low decoherence rates and therefore can provide a longer coherence time than the usually adopted harmonic trap. 

To understand atomic decoherence in an optical dipole trap with an arbitrary spatial form, we first present a theoretical analysis.  
We consider a one-dimensional (1D) trap with spatial form $V(x)= \lambda x^{2l}$ ($l$ is an integer), then the Hamiltonian of a particle in the trap is 
\begin{equation}
H=\frac{p^2}{2M} + \lambda x^{2l}, \label{H1}
\end{equation}
where $p$ is the momentum, $M$ is the mass of the particle.  
Assume a harmonic oscillator (HO) with oscillation frequency $\omega$ which fulfills the relation
\begin{equation}
\lambda=\frac{\hbar \omega}{4} \left( \frac{2 M \omega}{\hbar} \right)^{l}, \label{D1}
\end{equation}
the Hamiltonian (\ref{H1}) can be reduced to
\begin{equation}
H=\frac{\hbar \omega}{4} \left( P^2+X^{2l}\right), \label{H2}
\end{equation}
where $P=\sqrt{\frac{2}{\hbar M \omega}} p$ and $X=\sqrt{\frac{2 M \omega}{\hbar}} x$ are the dimensionless momentum and coordinate. 
For $l=1$, we have $\lambda=M \omega^2/2$, and the Hamiltonian reduces to the HO form. 
The quantum energy and wavefunction of the Hamiltonian (\ref{H1}) with $V(x)= \lambda x^{2l}$ can be obtained in the Hilbert space spanned by the eigen-wavefunction of the HO with oscillation frequency $\omega$. 

Next, we will analyze the variance of the DES and PJID of the electronic ground states in an optically trapped atom caused by the noise associated with the trap light.
For a real trap whose form is approximated by $V(x)= \lambda x^{2l}$, $\lambda$ usually can be expressed as $\lambda=V_c/a^{2l}$ with $V_c$ the characteristic trap potential and $a$ the characteristic size.
Therefore, the trap frequency for the corresponding HO would be 
\begin{equation}
\omega=\left(\frac{4 V_c}{\hbar}\right)^{\frac{1}{l+1}} \left(\frac{ \hbar}{2 M a^2} \right)^{\frac{l}{l+1}}. \label{frequency}
\end{equation}
If a thermal atom with temperature $T$ is considered, by following the procedures in \cite{PFYang_2022} the variance DES between the two electronic ground states of the trapped atom can be expressed as 
\begin{align}
    \text{Var} (\Delta^\text{DES})&= -\eta \frac{V_0}{\hbar} \frac{\text{Var} (P)}{\Bar{P}} +\frac{\eta}{l+1} \frac{k_B T}{\hbar} \frac{\text{Var} (P)}{\Bar{P}} \label{VarDFS}
\end{align}
In this equation, $\eta$ is a factor that reflects the influence of hyperfine splitting between the two electronic states on the DES, $V_0$ is the trap potential at the position of the trapped atom, $\Bar{P}_0$ is the average power of the trap light, $\text{Var} (P_0)$ is the variance of the power, and $k_B$ is the Boltzmann constant. 
We have $k_B T=\langle n\rangle \hbar \omega$ with $\langle n\rangle$ the corresponding mean phonon number of the HO.
If $V_0$ can be set as zero, e.g., in a well-aligned blue-detuned trap, the second term will dominate $\text{Var} (\Delta^\text{DES})$. Atoms trapped with a potential with higher $l$ will suffer a smaller decoherence due to the smaller $\text{Var} (\Delta^\text{DES})$ for a given relative power noise ${\text{Var} (P)}/{\Bar{P}}$.

For the decoherence caused by the PJID, we need to calculate the phonon jumping rate (PJR). The two heating sources, the heating due to the parametric noise and the point noise, can be dealt with separately in the perturbation theory \cite{Savard1997, Gehm1998}. The fractional fluctuations in $\lambda$ and $x$ are assumed to be $\epsilon_\lambda(t)$ and $\epsilon_x(t)$, respectively. The perturbations of interest are then
\begin{equation}
H'=\epsilon_\lambda(t) \lambda x^{2l}, \label{PH1}
\end{equation}
and
\begin{equation}
H'= 2 l \lambda x^{2l-1} \epsilon_x(t). \label{PH2}
\end{equation}
If the trapped atom is prepared in the phonon state $|n\rangle$ at time $t=0$, the average rate that it transits to another state $|m\ne n\rangle$ due to the two perturbations is 
\begin{equation}
R_{\lambda}= \frac{\pi \omega^2}{16} S_\lambda (\omega_{mn}) \sum_{m\ne n} |  \langle m | X^{2l}| n \rangle|^2 \label{rate1}
\end{equation}
and 
\begin{equation}
R_{x}= \frac{\pi}{2 \hbar} M \omega^3 S_x (\omega_{mn}) l^2 \sum_{m\ne n} |  \langle m | X^{2l-1}| n \rangle|^2, \label{rate2}
\end{equation}
respectively. Here, $S_\lambda(\omega)$ and $S_x(\omega)$ are the one-sided power spectrum of the fractional fluctuations in $\lambda$ and $x$ that are determined by the intensity and pointing noises of the optical trap beam; $\omega_{mn}=(E_m-E_n)/\hbar$ is the transition frequency between $|n\rangle$ and $|m\rangle$. $|\langle m | X^{2l-1}| n \rangle|^2$ and $|\langle m | X^{2l}| n \rangle|^2$ can be calculated in the Hilbert space spanned by the quantum states of the HO.

According to our previous work, the overall coherence can be written as
\begin{equation}  \label{coherence}
 \text{C}(t)=\textbf{e}^{-\left[\text{Var} (\Delta^\text{DES})\right]^2 t^2 /2- R t},
\end{equation} where $R= R_{\lambda} + R_{x}$ is the overall phonon jumping rate. In a three-dimensional trap, the PJR in all dimensions needs to be summed over to give the overall $R$.
From Eq. (\ref{VarDFS}) we can see that $\text{Var} (\Delta^\text{DES})$ becomes smaller for larger $l$.
In order to compare the PJR in optical dipole traps with different spatial forms, we make the following assumptions: 1) the traps have the same characteristic potential $V_c$ and size $a$; 2) the trapped atom is prepared in its ground phonon states; 3) the $S_\lambda(\omega)$ ($S_x(\omega)$ ) stays the same for different traps.
For a harmonic trap, we have $l=1$, $\sum_{m\ne 0} |  \langle m | X^{2l}| 0 \rangle|^2=2$, and $l^2 \sum_{m\ne 0} |  \langle m | X^{2l-1}| 0 \rangle|^2=1$.
For a quartic trap, we have $l=2$, $\sum_{m\ne 0} |  \langle m | X^{2l}| 0 \rangle|^2=5.83$, and $l^2 \sum_{m\ne 0} |  \langle m | X^{2l-1}| 0 \rangle|^2=8.48$.
Combined with Eqs. (\ref{rate1}) and (\ref{rate2}), we finally get $R_\lambda^{l=2}<R_\lambda^{l=1}$ under the condition ${\omega^{l=2}}/{\omega^{l=1}}<0.59$ and $R_x^{l=2}<R_x^{l=1}$ under the condition ${\omega^{l=2}}/{\omega^{l=1}}<0.49$. 
The smaller ${\omega^{l=2}}/{\omega^{l=1}}$ ratio, the lower the PJR in the quartic trap than in the harmonic counterpart.
The situation in which the ODT has even higher-order spatial form can be analyzed in the same way.

\begin{figure}[t]
\centering
\includegraphics[width=0.8 \columnwidth]{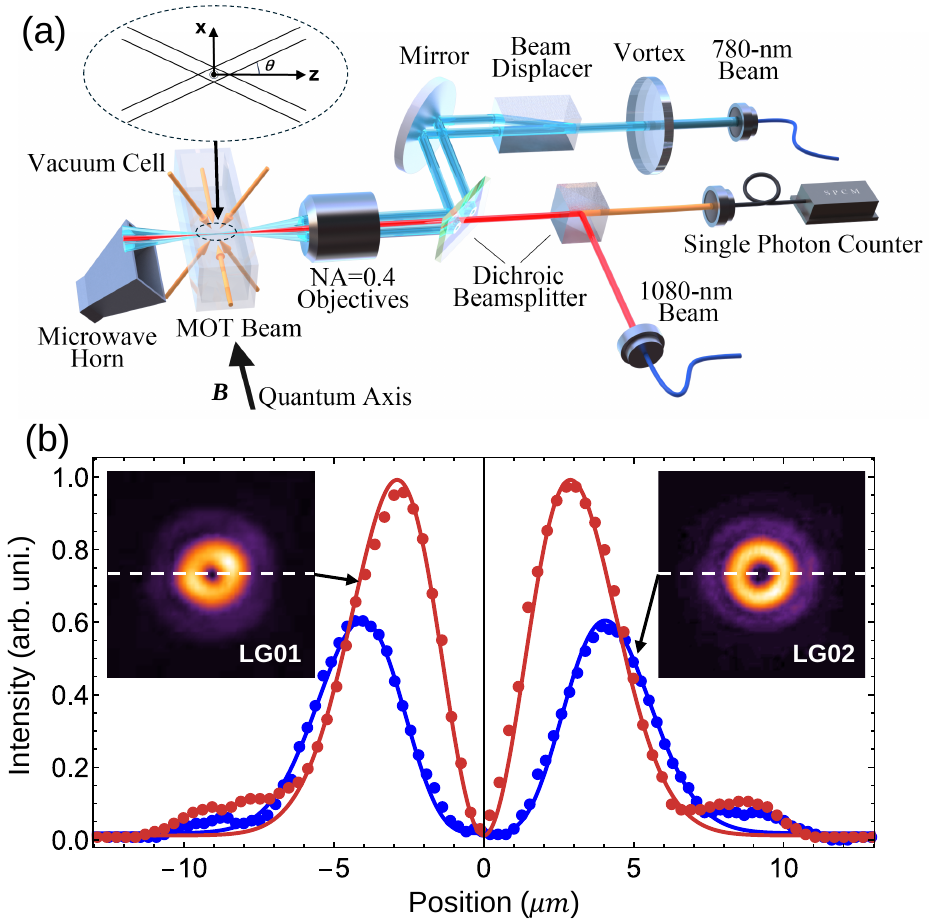}
\caption{(a) The experimental setup. Single Cs atom is captured by the 1080-nm RODT from MOT in the vacuum cell and adiabatically transferred into the 780-nm BBT. A microwave from the horn is used for coherent manipulating the single atom. The atomic fluorescence is collected by the objective and detected using a single-photon detector. The sketch in the dashed circle shows the orientation of the two crossing donut beam at the focus of the objective.
(b) The intensity distribution (solid dots) and the data fitting (solid lines) of LG01 and LG02 laser beams along the dashed lines in the inset images, which are the real images of LG01 and LG02 beams. The fitted waists of LG01 and LG02 beams are 4.09 and 4.05 $\mu$m, respectively. }
\label{setup}
\end{figure}

To verify the theory, we constructed two optical dipole traps with harmonic and quartic spatial forms and compared the coherence time of the trapped atoms. 
We adopted Laguerre-Gaussian (LG) laser beams with wavelength $\lambda=780$ nm to build blue-detuned bottle traps (BBTs) to trap cesium (Cs) atoms.
The experimental setup is similar to \cite{Li2012} and the schematic is demonstrated in Fig. \ref{setup}(a). 
A collimated 780-nm Gaussian-mode laser beam is converted into a hollow LG beam (also known as a donut beam) by vortex phase plates. Vortex phase plates with topological charge m=1 or 2 can create LG01 or LG02 mode beams. 
A beam displacer is then used to split the donut beam into two spatially parallel beams separated by 4 mm. The polarizations of these two beams are orthogonal to each other. Then, a homemade NA = 0.4 objective with a focal length f = 28.4 mm focuses these two donut beams.
The two donut beams cross with each other in the focus at an angle $2 \theta = 8^{\circ}$ and form a BBT.

To assess the mode quality of the LG beams, we captured images of these donut beams at equal power using a CCD camera. The insets in Fig. \ref{setup}(b) show the images for the LG01 and LG02 beams, respectively. The intensity distribution across the center [shown as the dashed lines on the images] is extracted and fitted by the formula
\begin{equation}
I\left(r\right)=\frac{P}{\pi w_{}^2} \left(C_{lp}^{LG}\right)^2\left(\frac{2r^2}{w^2}\right)^{|l|}\exp\left(-\frac{2r^2}{w^2}\right)\left(L_{p}^{|l|}\left(\frac{2r^2}{w^2}\right)\right)^2,
\label{LG}
\end{equation} 
where $P$ is the total light power, $C_{lp}^{LG}=\sqrt{\frac{2p!}{\pi(l+p)!}}$; $L_{p}^{|l|}$ is the Laguerre polynomial; $l$ and $p$ are the order of the orbital angular momentum and radial mode index; $w=w_0\sqrt{1+(z/z_R)^2}$ with $w_0$ the waist and $z_R=\pi w_0^2/\lambda$ the Rayleigh length of the LG beam. Figure \ref{setup} (b) shows the data and the fitting curves. We can see that the intensity barrier of the LG01 beam is higher than that of the LG02 beam. The height ratio is 1.59, which is higher than the ideal value of 1.36 due to the larger diffraction of the LG02 phase plate.
It is obvious that the spatial distribution around the core region of the beam is consistent with the theory.

\begin{figure}[t]
\centering
\includegraphics[width=\columnwidth]{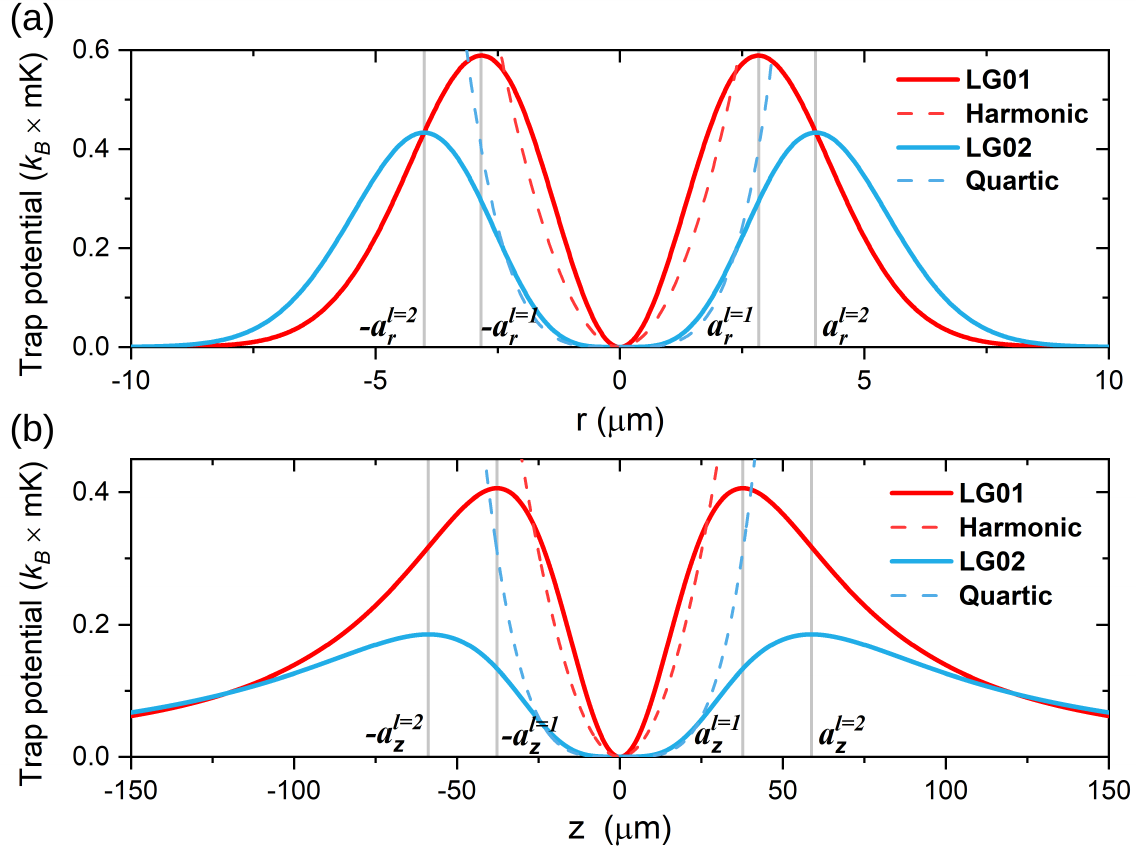}
\caption{The spatial dependence of the BBTs on $r$ (a) and $z$ (b). Due to the small angle $\theta$, the difference of the BBT spatial forms between $x$ and $y$ directions is tiny. Therefore, they are displayed by the same line with the horizontal coordinate $r$. The solid and dashed lines are the real potential distribution and the approximated harmonic trap. The red and blue lines represent the BBTs with LG01 and LG02 laser beams, respectively. The light gray vertical lines mark the bounds of the BBTs.
}
\label{fig2}
\end{figure}

The BBT with LG01 (LG02) beam can be approximated by the harmonic (quartic) trap.
The theoretical potential distributions along the coordinate axes of the BBT are shown in Fig. \ref{fig2}.
The expansions of the BBT potential with LG01 and LG02 beams around the trap center are \cite{Zhang2011}
\begin{align}
    V^{l=1}(x,0,0)&= V_0 \frac{x^2}{\left(\frac{w_0}{2 \cos{\theta}}\right)^2}+O(x^4) \label{HVx}\\ 
    V^{l=1}(0,y,0)&= V_0 \frac{y^2}{(w_0/2)^2}+O(y^4) \label{HVy}\\
    V^{l=1}(0,0,z)&= V_0 \frac{z^2}{\left(\frac{w_0}{2 \sin{\theta}}\right)^2}+O(z^4) \label{HVz}\
\end{align}
and 
\begin{align}
    V^{l=2}(x,0,0)&= V_0 \left[\frac{x^2}{\left(\frac{w_0}{\sqrt{2}\cos{\theta}}\right)^2} \right]^2+O(x^6) \label{QVx}\\\
    V^{l=2}(0,y,0)&= V_0 \left[\frac{y^2}{(w_0/\sqrt{2})^2}\right]^2+O(y^6) \label{QVy}\\
    V^{l=2}(0,0,z)&= V_0 \left[\frac{z^2}{\left(\frac{w_0}{\sqrt{2}\sin{\theta}}\right)^2}\right]^2+O(z^6) \label{QVz}
\end{align}
respectively. Here, $V_0= \frac{\alpha}{\pi} \frac{P}{\pi w_0^2}$ with $\alpha$ the polarizability can be defined as the characteristic trap potential. The characteristic trap sizes are $\{a_x,a_y,a_z\}^{l=1}=\{w_0/(2 \cos{\theta}),w_0/2,w_0/(2 \sin{\theta})\}$ and $\{a_x,a_y,a_z\}^{l=2}=\{w_0/(\sqrt{2} \cos{\theta}), w_0/\sqrt{2}, w_0/(\sqrt{2} \sin{\theta})\}$ for BBTs with LG01 and LG02 beams, respectively. The characteristic sizes are exactly the coordinates of the trap bounds, as shown in Fig. \ref{fig2}.
Higher-order terms can be omitted in real experiments. Therefore, the theory of atom decoherence in different traps can be checked with these BBTs. 

\begin{figure}[t]
\centering
\includegraphics[width=\columnwidth]{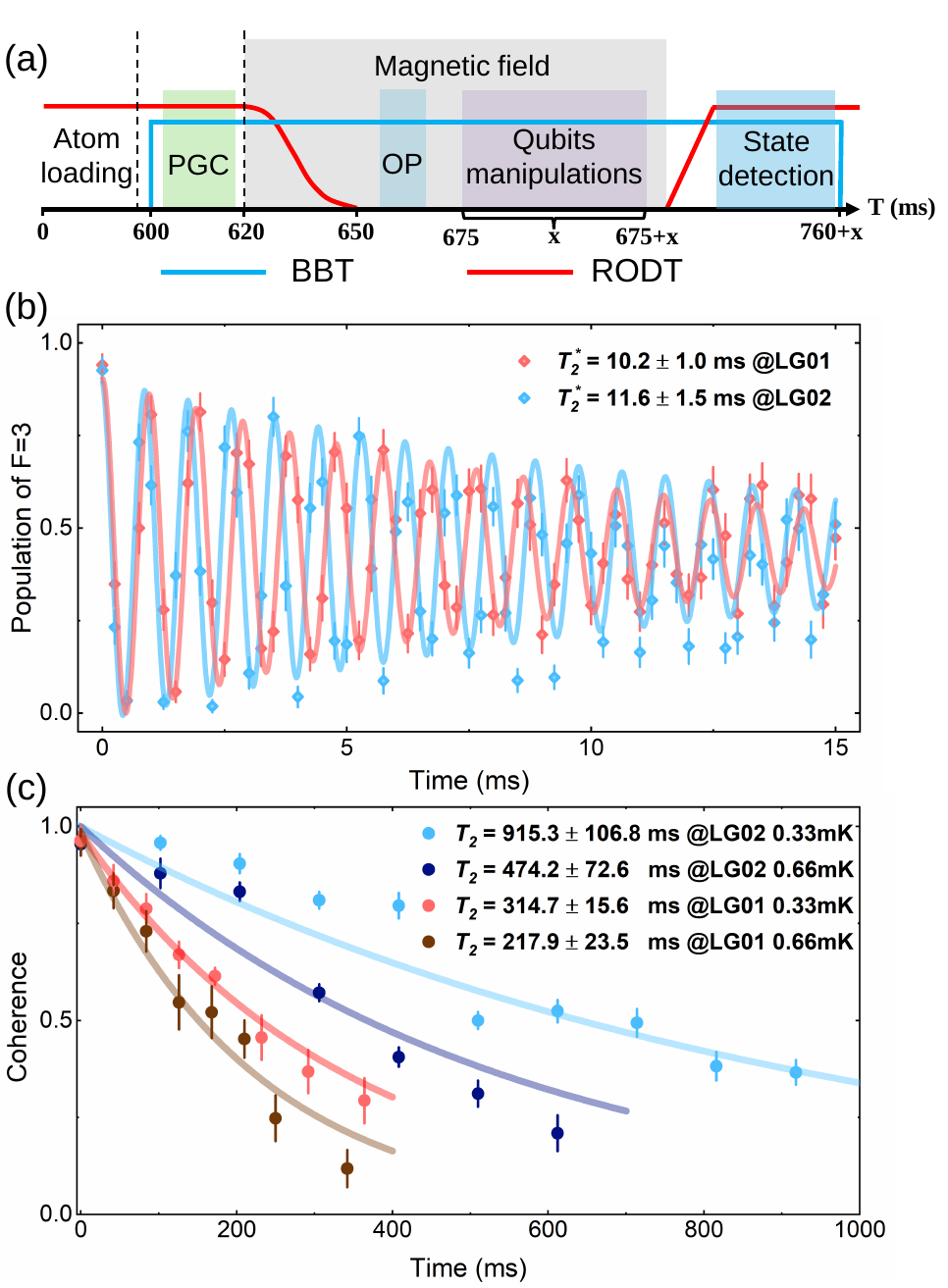}
\caption{Measurements of the coherence time. (a) The time sequence of the experiment. (b) The fringe of the Ramsey interferometer. (c) The decay of the coherence obtained by spin-echo under different trap settings.}
\label{fig3}
\end{figure}

To efficiently load the single atoms into the BBT, an auxiliary 1080 nm red-detuned optical dipole trap (RODT) with a waist of 2 $\mu m$ is employed to load a single cesium (Cs) atom from a cold Cs atomic ensemble. 
The center of the RODT is carefully aligned to overlap with the BBT. 
The fluorescence from the optically trapped single Cs atom, collected by the same objective, is sent to a single photon counter for detection. 
A 9.2-GHz microwave signal is applied by a microwave horn to coherently drive the Cs clock transition ($|F=4, m_F=0 \rangle \leftrightarrow |F=3, m_F=0 \rangle$). 
The time sequence for the main part of the experiment is shown in Fig. \ref{fig3} (a), where the preparation phase for the cold atom ensemble via magneto-optical trap (MOT) is omitted. 
This sequence begins with the loading of single Cs atoms from the MOT by RODT with a time tag of 0 ms. 
The loading process lasts for 500 ms, during which the collisional blockade mechanism guarantees that a single Cs atom is loaded in the RODT. 
At 600 ms, the BBT is switched on and the single Cs is held by the combination of RODT and BBT for 20 ms. 
During this time, a 10-ms polarization gradient cooling is applied to cool the temperature to 10 $\mu$K.
At 620 ms, the RODT is adiabatically switched off within 30 ms, leaving the BBT alone to hold the atom.  
It should be noted that the atom temperature drops to 5 $\mu$K after the adiabatic process.
A 3-ms optical pumping (OP) phase is followed to prepare the atom in the $|F=4, m_F=0 \rangle$ state.
At 675 ms, the microwave pulses are applied to coherently manipulate the qubit. 
Finally, the RODT is switched on and the atom state is read by a state detection phase.

According to theory \cite{2005diter, PFYang_2022}, a thermal atom trapped in an optical trap suffers mainly from inhomogeneous decoherence. 
The coherence time is mainly determined by the thermal temperature.
The decoherence process can be measured using the standard Ramsey interferometer, in which two $\pi/2$-pulses are applied to the atom with a variable time delay in between.
We studied this inhomogeneous decoherence process in both traps with the same trap potential at $V_0=k_B \times 0.33$ mK. 
The results are shown in Fig. \ref{fig2}(b).
The inhomogeneous coherence time $T^*_2$ of the atom in LG01 and the LG02 BBTs are $10.2\pm1.0$ and $11.6\pm1.5$ ms, respectively. 
Inhomogeneous coherence times are almost the same, which means the same temperature.
The dependence of $T^*_2$ on atom temperature given in \cite{2005diter} is
\begin{equation}
T_2^*=0.97\frac{2\hbar}{\eta k_BT}, \label{T2*}
\end{equation}
where $k_B$ is the Boltzmann constant. 
Therefore, we can infer that the atom temperatures in LG01 and LG02 BBTs are 5.8 $\mu$K and 5.1 $\mu$K, respectively. 
The temperatures are consistent with the results from the release-and-recapture method.

The inhomogeneous decoherence of the atom due to the thermal motion can be recovered by the dynamic decoupling techniques. 
Here, we adopted the simplest spin-echo process, in which a single $\pi$-pulse is inserted between two $\pi/2$-pulses, to recover the inhomogeneous decoherence.
The residual decoherence is homogeneous and mainly determined by DES and PJID \cite{Tian2024}.
To compare homogeneous decoherence in the two types of traps, we measured the decay of coherence with time for the atom in traps with characteristic barrier heights of $V_0=k_B \times 0.33$ mK and $V_0=k_B \times 0.66$ mK, separately.
The results are shown in Fig. \ref{fig3}(c).

We can see that the coherence time of a single atom in LG02 BBT is longer than that in LG01 BBT for the same $V_0$. However, because the characteristic trap sizes are different, direct comparisons are of little significance.
Fortunately, if we decrease the characteristic trap potential of LG01 BBT from $V_0$ to $V_0/2$, Eq. (\ref{HVx}) can be rewritten as 
\begin{align}
    V^{l=1}(x,0,0)&= V_0 \frac{x^2}{\left(\frac{w_0}{\sqrt{2} \cos{\theta}}\right)^2}+O(x^4). \label{HVx1}
\end{align}
Eqs. (\ref{HVy}) and (\ref{HVz}) can be rewritten similarly.
This means that the LG01 BBT with characteristic trap potential $V_0/2$ and sizes $\{a_x,a_y,a_z\}^{l=1}=\{w_0/(2 \cos{\theta}),w_0/2,w_0/(2 \sin{\theta})\}$ is equivalent to that with potential $V_0$ and sizes $\{a_x,a_y,a_z\}^{l=1}=\{w_0/(\sqrt{2} \cos{\theta}), w_0/\sqrt{2}, w_0/(\sqrt{2} \sin{\theta})\}$.
Thus, the comparison of atom coherence time in LG01 BBT with $V_0=k_B \times 0.33$ mK to that in LG02 BBT with $V_0=k_B \times 0.66$ mK makes more sense.
From Eq. (\ref{frequency}) we have ${\omega^{l=2}}/{\omega^{l=1}}=\left(\frac{\hbar^2}{8M a^2 V_0}\right)^\frac{1}{6}=0.074$. Combined with Eq. (\ref{VarDFS}), we can conclude that the BBT with LG02 beams has a lower decoherence rate than that with LG01 beams in theory.
The experimental results of the coherence time $474.2\pm72.6$ ms in LG02 BBT versus $314.7\pm15.6$ ms in LG01 BBT solidly support the theoretical expectation.
These results qualitatively verify our proposal that the ODT with a higher-order spatial form possesses a lower decoherence mechanism.

In conclusion, we proposed that the ODT with higher-order spatial form can provide an intrinsic lower decoherence rate than the counterpart with lower-order spatial form.
We first formulated the atom decoherence rate by analyzing the variance of DES and the PJR, which are known as the two main decoherence mechanisms.
Then, we experimentally proved our proposal by constructing blue-detuned harmonic and quartic traps with LG01 and LG02 laser beams and comparing the atom coherence time.
Our method offers an alternative approach to enhancing atomic coherence time by modifying the spatial form of the ODT.

\begin{backmatter}
\bmsection{Funding} This work was supported by National Key Research and Development Program of China (2021YFA1402002); Innovation Program for Quantum Science and Technology (Grant No. 2023ZD0300400); National Natural Science Foundation of China (U21A6006, U21A20433, 92465201, 12474360, and 92265108).

\bmsection{Disclosures} The authors declare no conflicts of interest.

\bmsection{Data availability} Data underlying the results presented in this paper are not publicly available at this time but may be obtained from the authors upon reasonable request.
\end{backmatter}

% Bibliography

% Full bibliography added automatically for Optics Letters submissions; the following line will simply be ignored if submitting to other journals.
% Note that this extra page will not count against page length
%\bibliographyfullrefs{sample}

%Manual citation list
%\begin{thebibliography}{1}
%\bibitem{Zhang:14}
%Y.~Zhang, S.~Qiao, L.~Sun, Q.~W. Shi, W.~Huang, %L.~Li, and Z.~Yang,
 % \enquote{Photoinduced active terahertz metamaterials with nanostructured
  %vanadium dioxide film deposited by sol-gel method,} Opt. Express \textbf{22},
  %11070--11078 (2014).
%\end{thebibliography}

% Please include bios and photos of all authors for aop articles
\ifthenelse{\equal{\journalref}{aop}}{%
\section*{Author Biographies}
\begingroup
\setlength\intextsep{0pt}
\begin{minipage}[t][6.3cm][t]{1.0\textwidth} % Adjust height [6.3cm] as required for separation of bio photos.
  \begin{wrapfigure}{L}{0.25\textwidth}
    \includegraphics[width=0.25\textwidth]{john_smith.eps}
  \end{wrapfigure}
  \noindent
  {\bfseries John Smith} received his BSc (Mathematics) in 2000 from The University of Maryland. His research interests include lasers and optics.
\end{minipage}
\begin{minipage}{1.0\textwidth}
  \begin{wrapfigure}{L}{0.25\textwidth}
    \includegraphics[width=0.25\textwidth]{alice_smith.eps}
  \end{wrapfigure}
  \noindent
  {\bfseries Alice Smith} also received her BSc (Mathematics) in 2000 from The University of Maryland. Her research interests also include lasers and optics.
\end{minipage}
\endgroup
}{}

\bibliography{sample}
\bibliographyfullrefs{sample}
\end{document}